# BIG IS FRAGILE: AN ATTEMPT AT THEORIZING SCALE

Atif Ansar[1,*], Bent Flyvbjerg[1], Alexander Budzier[1], Daniel Lunn[2]

**Affiliations:**
[1]Saïd Business School, University of Oxford, OX1 1HP, UK.
[2]Department of Statistics, University of Oxford, OX1 3GT, UK.

* To whom correspondence should be addressed.
E-mail: atif.ansar@sbs.ox.ac.uk

> *"There is nothing more deceptive than an obvious fact."*
> Sherlock Holmes in *The Boscombe Valley Mystery*, Sir Arthur Conan Doyle.

# INTRODUCTION

Theories of big have advocated the proposition that "bigger is better" since the mid-nineteenth century, drawing especially on notions of economies of scale and scope (Stigler, 1960; Silberston, 1972 Chandler, 1990), natural monopoly (Mill, 1848; Mosca, 2008), or preemptive capacity building (Porter, 1980). Building big has traditionally been seen as necessary to secure efficient economies of scale and to lock competitors out of future rivalry (Ghemawat, 1991; Ghemawat & der Sol, 1998; Hayes & Garvin, 1982, p. 78; Penrose, 1959, 2009, pp. 89-92; Wernerfelt & Karnani, 1987). Economies of scale in the production of electricity, for example, assume declining average cost curve as output expands (Ansar & Pohlers, 2014: Figure 1).

Theories of big have held an enduring sway on management practice. Scaling is deemed necessary for survival and competitive advantage in many industries such as steel (Crompton & Lesourd, 2008), shipping (Cullinane & Khanna, 2000), banking (Cavallo & Rossi, 2001), telecommunications (Majumdar & Venkataraman, 1998; Foreman & Beauvais, 1999), water, and electricity utilities (Kowka, 2002; Pollitt & Steer, 2011). Big ventures are witnessed when companies try to pursue new product or geographic markets: Motorola spent US$ 6 billion in technological rollout of its Iridium satellite constellation (Bhidé, 2000); the German steel company ThyssenKrupp spent over US$ 11 billion to enter the US and Brazilian markets (Ansar, 2012). In the context of economic development, Sachs (2006) has popularized the view that big challenges—such as poverty alleviation, energy and water scarcity, or urbanization can only be solved with "big push" solutions.

Theories of big are not universally accepted, needless to say. Schumacher (1973) championed, "Small is beautiful." Easterly (2002, 2006) has argued against the big push "mega-reforms" that Sachs favors. Perhaps the most persistent attack on theories of big has come from Lindblom who advanced the notions of "muddling through" (1959) and "disjointed incrementalism" (1979) as an alternative to "bigger is better." In the academic debate, a school of thought emerged around the notion of incrementalism, with concepts such as logical incrementalism (Quinn, 1978); modularity (Baldwin & Clark, 2000); real optionality (Copeland & Tufano, 2004); adaptive change (Heifetz et al., 2009). The key rationale for incrementalism is based on the well-documented limits of rationality in decision making under uncertainty (Atkinson, 2011, p. 9).

Yet theories of big have maintained an enduring upper hand in mainstream business and government practice: megaprojects, justified on the bases of theories of big, are more ubiquitous and bigger than ever (Flyvbjerg, 2014). A renewed enquiry into the validity of theories of big has become urgent because big capital investment decisions routinely fail in the real world. Motorola's Iridium went bankrupt within three years of its launch wiping nearly US$10 billion of equity (Kerzner, 2009: 351). ThyssenKrupp's megaprojects in Americas initiated in 2005 suffered a worse fate. By June 2012 ThyssenKrupp's survival was in doubt and the company's share price was a quarter of its May 2008 high. The company was forced to raise fresh capital in part by selling its ill-fated US steel plant at a steep discount to its archrivals ArcelorMittal and Nippon-Sumitomo. ThyssenKrupp's top management team, including the Chairman Gerhard Cromme, who initiated the big capital investments, was fired. Other cases of big bets gone awry abound. Losses from Boston's Big Dig or Citigroup's "bold bets" reach into the billions of dollars (Dash & Creswell, 2008). Not only did the costs sunk into Japan's Fukushima nuclear power plant become unrecoverable; its cleanup costs rippled far into the economy. Beyond the financial calculus, even the reticent Chinese government is opening up to the profound and unforeseen human and environmental impacts of China's Three Gorges dam (Qiu, 2011; Stone, 2008, 2011).

Over the last fifteen years evidence from large datasets has mounted that the financial, social and environmental performance of big capital investments, in the public and private sectors alike, is strikingly poor (see, for example, Nutt, 1999, 2002; Flyvbjerg et al, 2002, 2003, 2005, 2009; Titman, Wei, and Xie, 2004; Flyvbjerg & Budzier, 2011; Van Oorschot et al., 2013; Ansar et al., 2014).

In this chapter we characterise the propensity of big capital investments to systematically deliver poor outcomes as "fragility" a notion suggested by Taleb (2012). A thing or system that is easily harmed by randomness is fragile. We argue that, contrary to their appearance, big capital investments break easily—i.e. deliver negative net present value—due to various sources of uncertainty that impact them during their long gestation, implementation, and operation. We do not refute the existence of economies of scale and scope. Instead we argue that big capital investments have a disproportionate (non-linear) exposure to uncertainties that deliver poor or negative returns above and beyond their economies of scale and scope. We further argue that to succeed, leaders of capital projects need to carefully consider where scaling pays off and where it does not. To automatically assume that "bigger is better," which is common in megaproject management, is a recipe for failure.

Below, first we develop and draw linkages between three theoretical concepts: Big, Fragility, and Scale versus Scalability. Building on the relevant literature, we advance twelve propositions about the sources and consequences of fragility, which we then apply to big capital investments. Specifically, we develop the construct of "investment fragility"—the threshold at which the cumulative losses from an investment exceeds the investment's cumulative gains. Second, we substantiate our propositions about the fragility of big capital investments using evidence from a dataset of 245 big dams. Our theoretical propositions and evidence from big dams suggests that big capital investments fail to achieve the aspirations of scalability and efficiency set for them at the outset. Once broken, these investments become stranded causing long-term harm to the companies and countries that undertake them.

# THEORY AND CONSTRUCTS

**What Is Big?**
Whether something is big seems obvious enough. Yet the construct of big and the related ideas of large, large-scale, major, mega, or huge etc. prove deceptively elusive when subjected to scrutiny. A literature review of definitions of big and related words reveals a non-exhaustive list of multiple dimensions of the construct of big such as:

- Physical proportions measured in height, length, mass, weight, area, or volume;
- Inputs required to build and run the thing measured in terms of quantities (and quality) of land, labor, or equipment required;
- Financial outlay measured in upfront capital expenditure (Flyvbjerg, 2014), recurrent operational expenditure, or end-of-life costs (Clark & Wrigley, 1995, 1997);
- Supply measured in the units of output that the thing can produce (Samuelson, 1948; Stigler, 1958) or the multiplicity of outputs (Chandler, 1990);
- Demand being served measured not only in terms of units of demand but also the quality, speed, and functionality (Weinstock & Goodenough, 2006);
- Temporality measured in time it takes to build the thing or the length of its life span (Gomez-Ibanez, 2003);
- Spatial fixity or immobility and the cost of moving an asset—big tends to "spatial fixity" (Ansar, 2012);
- Complexity measured, for example, with a focus on the technical aspects of the asset or its delivery (Simon, 1962); or with a focus on the social and political complexity (Liu et al., 2007);
- Impact measured in number of people who might benefit or be harmed; number of functions enabled or disabled; or the magnitude and pace of change the thing can cause in its environment.

The relationships among these multiple dimensions—and even the indicators within each dimension—are seldom straightforward. For instance, it is often taken for granted that in order to reduce the amount of time required to complete a venture, e.g. a software IT project, a manager would need to mobilize more programmers, which will also increase the capital expenditure of the venture (for a discussion see Atkinson, 1999; Williams, 2005).

Bigness entails multiple and unpredictable interactions across the dimensions we have listed with which theories of big—such as the notion of economies of

scale—have not meaningfully engaged. The risk of big is reflected in these interactions, which we turn to now.

**What Is Fragility?**
Taleb (2012) proposes the construct of fragility, and its antonym antifragile, to capture the type of randomness and risk we talk about here. Taleb's inspiration for the concept came from his observation that there is no random event that can benefit a porcelain cup resting on a table yet a wide range of uncertain events – such as earthquakes or human clumsiness – that can pose harm (Taleb 2012, p. 268). Fragile things, like the porcelain cup, are vulnerable to members of the "disorder family" such as randomness, uncertainty, volatility, variability, disturbance, or entropy—terms we use interchangeably below[i] to explore the fragility of big capital investments.

Fragility as a construct has found increasing use in diverse fields such as construction (Shinozuka et al., 2000; Choi et al., 2004); human bones (Seref-Ferlengez et al., 2015); global financial and banking crises (Davis et al., 1995; Taleb et al., 2012; Klemkosky, 2013; Calomiris & Haber, 2014); mathematics (Taleb & Douady, 2013); ecology and ecosystems (Nilsson & Grelsson, 1995; Sole & Montoya, 2001); interdependent networks (Callaway et al., 2000; Buldyrev et al., 2010; Vespignani, 2010; Parshani et al., 2011; Gao et al., 2012; Morris & Barthelemy, 2013); material sciences, e.g. in assessing properties of glass forming liquids (Scopigno et al., 2001; Mauro et al, 2014; Martinez-Garcia et al., 2015); conflict-prone countries (USAID, 2005; Ziaja, 2012); democratic political systems (Mitchell, 1995; Issacharoff, 2007); and even personal ethics (Nussbaum, 2001). Despite its diverse uses, fragility has certain common meanings across these fields. Taking our point of departure in Taleb's work, because this is the most comprehensive, and supplementing this with other sources, we first distill a general definition of fragility. Second, we use this to develop a more specific concept called "investment fragility," which we apply to big capital investments.

Fragility describes, "how [a] system *suffers*" when it encounters disorder (Taleb and Douady, 2012, p. 1677, *italics in the original*). The outcome of fragility is typically an irreversible loss of functionality. Our preliminary proposition, which we'll refine below, is as follows:

> *Proposition 1. Fragility is typically irreversible. Once broken the fragile cannot be readily restored to its original function.*

Fragility as a material property is signified by a material's breaking point (Scopigno et al., 2003; Martinez-Garcia et al., 2015). Mauro et al. (2014) call this unique intrinsic fragility threshold a "structural signature". Buldyrev et al. (2010) conducted similar structural analysis on networks such as electricity grids (also see Vespignani, 2010). They find that intrinsic structural features of networks—such as number of nodes; whether a network is isolated or interdependent; and number of interconnections with other networks—can help determine the "critical threshold" at which the various networks break down[ii]. As a more generalizable proposition:

> *Proposition 2. A thing (material, system, process, or network) has a unique "fragility threshold", which careful structural analysis and experimentation can reveal, at which it will break down under influence from stressor(s). All other things being equal, the lower the preset threshold the greater the fragility.*

An assumed benefit of big is that the fragility threshold increases with size. However, Taleb (2012, pp. 278-80) argues that larger organisms such as elephants tend to have a lower fragility threshold as a proportion of their size than smaller organisms. Thus, for example, it might take a stone twice the body weight of a cat to cause a fatal outcome. But a boulder only half the bodyweight of an elephant might suffice for its fragility threshold. The source of this disproportionally greater fragility of bigger systems is to be found in Anderson's (XXXX) notion of "more is different." As a more generalizable proposition, we advance:

> *Proposition 3. As a system grows bigger, the relative size of a stressor required to break it will decline disproportionately.*

As suggested before another source of fragility is the interconnectedness of systems. In other words "inherited fragility." Inherited fragility helps to map out the diffusion of harm in the event of fragility in a corner of an interconnected system of systems. As a generalizable proposition:

> *Proposition 4. In an interdependent system of systems with no redundancy the threshold of fragility for the whole system of systems is the same as the threshold of the system's weakest component.*
> *Systems with a low fragility threshold and low function recoverability (Quadrant 4) require a great deal of cushioning to protect them from breaking.*

below suggests a composite way to map different things on the spectrum of greater to lesser fragility. This constitutes two dimensions: First, on the vertical axis is how easily something breaks, i.e. a high or low preset "fragility threshold." Second, on the horizontal axis is how easily can the broken thing be recovered to its self-same functional state once the fragility threshold is crossed, i.e. high or low "function recoverability."

However, not all things are as unlucky as Humpty the egg. A multi-use electric fuse also breaks easily but can also be easily restored to perform its sacrificial function (Quadrant 3). In contrast, a diamond has a high fragility threshold but once broken the damage is irreversible (Quadrant 2). Finally, fungi—and many systems in nature (Taleb, 2012)—do not break easily and easily recover (Quadrant 1). Incidentally, as our discussion on scalability in the next section shows, systems in Quadrant 1 share many properties with highly scalable systems[iii].

**Figure 1: A Map Of Fragility**

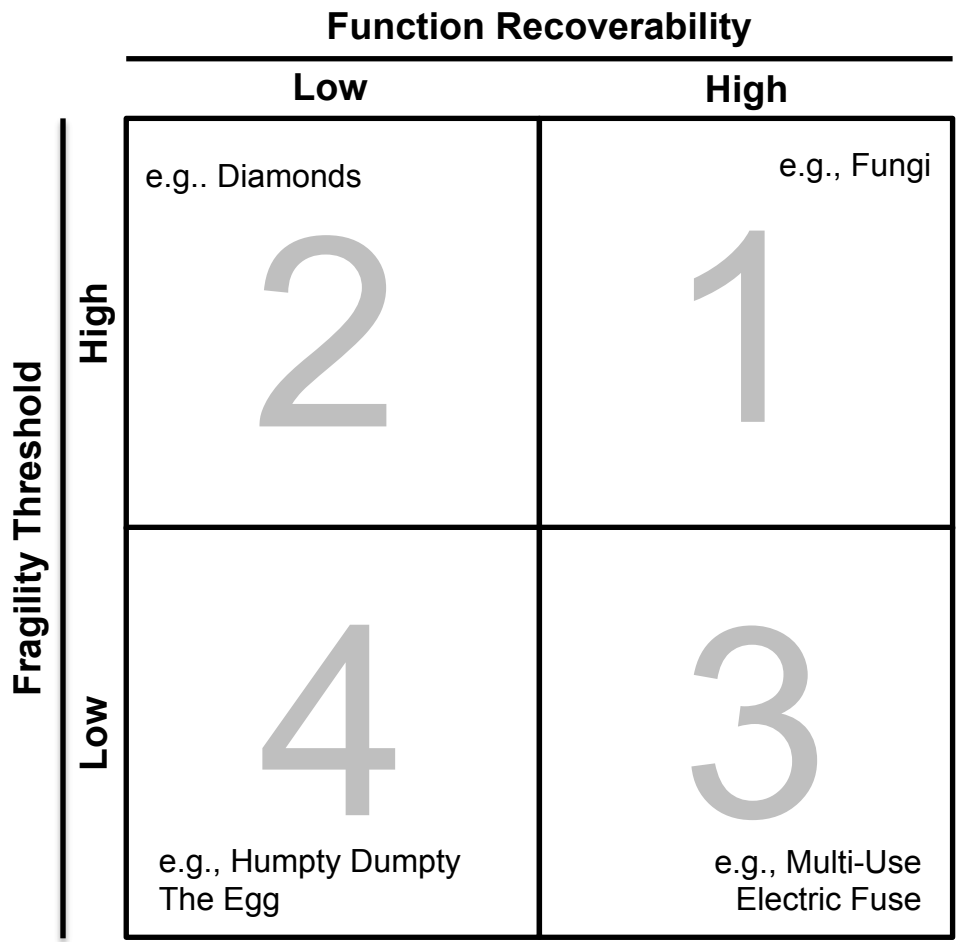

*Source: Authors*

> *Proposition 5. Systems with a low fragility threshold and low function recoverability (Quadrant 4) require a great deal of cushioning to protect them from breaking.*

Closely related to the distinction between "intrinsic" and "inherited" fragility is the distinction between "apparent" and "hidden" fragility. The egg's vulnerability to breaking is apparent. We thus handle eggs with care when bought at the grocery to cook for tomorrow's breakfast. We also do not act surprised when eggs break. Similarly, an electric fuse's fragility is apparent and in fact intended. In contrast, hidden vulnerability is insidious and entails surprise. Vespignani (2010, p. 985) finds that in cases of highly interconnected networks breakage is typically "abrupt" at a relatively low critical threshold, which introduces an element of surprise: "This makes complete system breakdown even more difficult to anticipate or control than in an isolated network."

Hidden fragility has an interaction effect with inherited fragility: Achilles' heel is a reminder of this phenomenon when a small hidden weakness is inherited by the otherwise mighty system[iv].

> *Proposition 6. The information conditions necessary to make a system truly robust are very strong because some minor source of hidden fragility is likely to remain. In instances of hidden and inherited fragility, seemingly robust systems will break down abruptly at a surprisingly low critical threshold. Lack of preparedness in instances of hidden and inherited fragility will accentuate the consequent harm.*

Although Taleb (2012, see, e.g., pp. 270-71) underplays the role of cumulative processes in creating fragility, hidden fragility is often created or exacerbated by them. Taleb typically focuses on fragility caused by one single blow. "Cumulative fragility," in contrast, is death by a thousand cuts. Keil & Montealegre (2000) liken such an ill-fated investment to a frog being boiled to death without knowing it (also see Keil, 1995; Royer, 2003).

Poorly understood dynamic changes often lower the fragility threshold without warning. Even a minor stressor, in such circumstances, leads to unexpected breakage. Seref-Ferlengez et al. (2015), for instance, find that human bones are constantly exposed to microdamage from routine stressors. Microdamage from routine wear and tear begins to accumulate that contributes "additively to reduced fracture toughness"—i.e. lowering of the threshold at which the bone breaks (Seref-Ferlengez et al., 2015, p.1). In technological systems accumulation of small errors is often behind man-made disasters (Turner and Pidgeon, 1997).

> *Proposition 7. Cumulative hidden processes will increase the propensity of fragility of a system over time.*

Hidden fragility also has sociological and psychological features. Scholars of power relationships suggest that organizations can tacitly choose to ignore the obvious (Flyvbjerg, 1998) or people in organizations "know that we do not *want* to know" (Ansar, 2015). Psychological biases such as overconfidence also "bring fragilities" due to "expert problems (in which the expert knows a lot but less than he thinks he does)" (Taleb, 2012, p. 215). Psychological theories predict that hidden fragilities in systems designed by experts will be the norm not the exception (Kahneman, 2011). NASA lost its $125 million Mars Climate Orbiter in September 1999 because one engineering team used metric units while another used imperial units for a spacecraft maneuver (Reichhardt, 1999; also see Hodgkinson & Starbuck, 2012, pp. 2-5 for discussion of NASA's space shuttle disasters in light of theories of organizational decision making).

"Our inability to recognize and correct this simple error has had major implications," said Edward Stone, the Director of the Jet Propulsion Laboratory (JPL) at NASA that oversaw the mission, according to CNN (30 September 1999). Also see Proposition 6 on information conditions of correcting hidden fragility.

> *Proposition 8. Behavioral and psychological biases will make the detection of hidden fragility more difficult. Hidden fragilities in systems designed by experts will be the norm not the exception.*

> *Proposition 9. In sociological and organizational contexts, hidden fragility will be hidden in plain sight.*

Although Taleb has tended to focus on external causes of disorder, ecologists working on fragility pay equal attention to internal changes that cause fragility (see Nilsson & Grelsson, 1995). Theorists of "broken symmetry" have also noted that as internal complexity increases, the threat of internal sources of disturbance—often driven by entirely random processes—grows non-linearly exposing a larger system to breakdown with no apparent cause (Anderson, 2011). From a consequentialist perspective, the distinction between external and internal stressors is immaterial: it is the magnitude of the negative outcome that follows that matters. Taleb (2012, p. 136) argues:

> You can control fragility a lot more than you think…detecting (anti)fragility—or, actually, smelling it[…]—is easier, much easier, than prediction and understanding the dynamics of events, the entire mission reduces to the central principle of what to do to minimize harm (and maximize gain) from forecasting errors, that is, to have things that don't fall apart, or even benefit, when we make a mistake...Not seeing a tsunami or an economic event coming is excusable; building something fragile to them is not.

From a consequentialist perspective, a generalizable proposition that can be advanced is as follows:

> *Proposition 10. If a system or process is systematically delivering poor outcomes, it is an indicator of fragility. Discard or redesign the system.*

> *Proposition 11. It is easier to be robust against the magnitude of harm that fragility might bring than predict or control the sources of stressor(s).*

**The Special Case Of "Investment Fragility"**
Above we analyzed general features of fragility. Now we turn to the special case of "investment fragility," which we define as *the vulnerability of a financial investment to becoming non-viable, i.e., losing its ability to create net economic value*. When applied to physical assets such as roads, oil refineries, bridges, factories, real estate developments etc., investment fragility causes them to become "stranded assets"—i.e. where "assets suffer from unanticipated or premature write-offs, downward revaluations or are converted to liabilities" (Ansar et al., 2013, p.2). A broken or stranded asset is a net drag on the economy—i.e. it consumes resources inefficiently that could have been put to alternative uses. An economy with too many assets prone to fragility is at a heightened risk of system-wide failure due to the domino-like effect of inherited fragility that can spread from one corner of the system of systems to the whole.

*Benefit-to-Cost Ratio (BCR): A Rule-Of-Thumb To Detect The Investment Fragility Threshold*
The fragility threshold of an investment can be located by modeling the investment's payoff structure. Figure 2 illustrates such a payoff structure graphically: the discounted stream of gain and the discounted stream of pain of a hypothetical investment is sorted in a descending order and then graphed. Since investments typically span long time horizons it is imperative to incorporate time value of money in the payoff structure by introducing an appropriate discount rate (see Souder & Shaver, 2010). The point of fragility in Figure 2 is where the cumulative pain—area "1" in Figure 2—just exceeds the cumulative gain—the area "2" in Figure 2. *In other words, an investment is broken when the cumulative gain-to-pain ratio falls below 1, less gain in the numerator than the pain in the denominator.* The gain-to-pain ratio, which is commonly thought of as benefit-to-cost ratio (BCR) in literature on capital budgeting, thus serves as a useful rule-of-thumb for detecting the fragility threshold of an investment.

Like the more general case of fragility, an asset suffers from greater *intrinsic* investment fragility when the absolute threshold of disturbance required to "break" it is low—e.g., a low *ex ante* benefit-to-cost ratio—whereby a stressor easily pushes the benefit-to-cost ratio below 1. The stressor may be, for example, just one weather event in the 60-year life of the asset that causes so much mess that costs of dealing with it spiral out of proportion. As with many fragile things, there is an element of surprise in investment fragility as well—seemingly improbable, and often minor, stressor(s) can cause disproportionate harm.

An investment is more fragile when once broken it is difficult or nearly impossible to reposition it to make recovery possible via alternative uses. Less specific assets—for example, buildings that can change use—are less prone to becoming irreversibly broken than assets with a higher specificity (see

Gómez-Ibáñez, 2003 for a rich ground of interconnections between investment fragility and discussion on "hold up" in transaction cost economics).

**Figure 2: Graphing "Investment Fragility"**

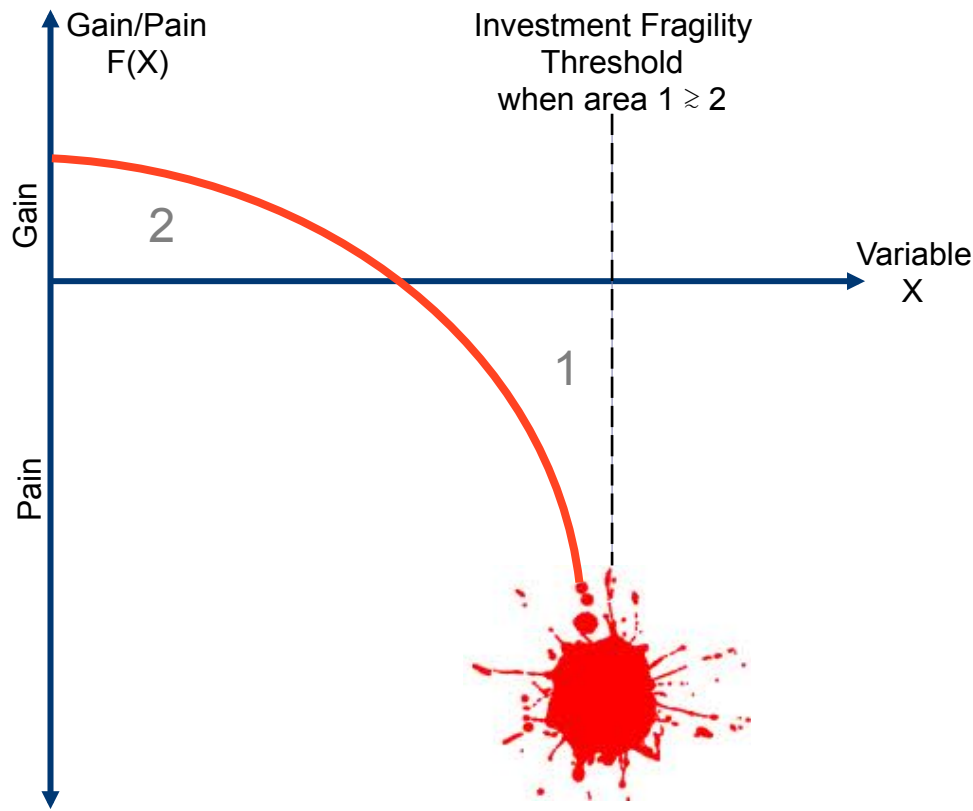

*Source: Adapted from Taleb (2012, p. 273).*

Similarly, an asset has higher intrinsic investment fragility when its gain (anticipated benefits and any windfall) is capped but the pain (anticipated cost and any unforeseen losses) is uncapped. This propensity exhibits itself in the shape of Figure 2 in where the top left part of the curve (gain) is bounded and finite but the bottom right part of the curve (pain) is unbounded and potentially an infinite pit. Based on this observation, Taleb (2012, p. 158, *italics in the original*) proposed: *"Fragility implies more to lose than to gain, equals more downside than upside, equals (unfavourable) asymmetry."*

**What Is Scalability?**
The Oxford English Dictionary defines "scalable" and "scalability" in the following manner:

"**scalable**, *adj.*
Able to be changed in scale. *rare.*
1977   *Jrnl. Royal Soc. Arts* 125 770/1   Such lasers are scalable since large volumes could be pumped uniformly."

"**scalaˈbility** *n.* the property of being scalable.
1978   *Sci. Amer.* Nov. 44/2   It took demonstrations of the scalability of the technology and tests of improved beam focusing...to catalyze an effort that led to support by the AEC."

We note, in particular, the unsatisfactory definition of the word scalable. What does the ability to be changed (or equally change without external force) in scale mean? In practice, scalability is most commonly discussed in the field of computer science, system architecture, and software programming (Hill, 1990; Gunther, 2007). This is insufficient for our purposes.

In academic literature, an understanding of whether, and under what conditions, something has the ability to be changed in scale has been deeply informed by advances in mathematics and particularly the field of fractal geometry. Mandelbrot's research identified two primary dimensions of scale: temporal and spatial. A grain of sand, a pebble, a rock, a cliff, and the coastline of say Britain represent a continuum of finer to coarser grades of the spatial scale (see Mandelbrot, 1967). Similarly, the price movements of a stock price over a scale between five seconds to five years represents finer to coarser grades of a temporal scale. Finer and coarser grades of scale can also be thought of in terms of degrees of zoom-in (microscopic) and zoom-out (macroscopic). To Mandelbrot's spatial and temporal scale, we introduce a third scale: "relational." The relational scale not only refers to the number of end-users (e.g., few or many) but also to the heterogeneity of end-users and their tailored need (see Ansar, 2010; Ansar, 2012 for a more extensive discussion on a *spatial*, *temporal*, and *relational* multi-scalar framework)[v].

Based on this understanding of scale, we define scalability as the ability of a thing (e.g., a system, system of systems, process, or network) to *effortlessly* transition back *and* forth from the very micro to the very macro *spatial*, *temporal*, and *relational* scales. Effortlessness connotes minimum friction in terms of the time, cost, etc. it takes to build up or remove capacity. Over longer time scales, effortlessness implies ability to quickly upgrade without losing compatibility. However, if scalability is understood and practiced as mainly the ability to scale up, with scant attention to scaling down – which is the dominant approach today in both the academy and practice (Sutton and Rao 2014) – then this in and of itself adds fragility. Here we therefore understand scalability as the ability to change in both directions, i.e., both up and down.[vi]

In contrasting big and scale we arrive at a key insight. Big typically possesses a degree of slack, which Weinstock and Goodenough (2006), call the "ability to handle increased workload (without adding resources to a system)." A big power plant, for example, rarely operates at full capacity. If demand were to increase from one segment of the day to another the power plant can be ramped up to meet some or all of the added demand. However, this limited slack is different from true scalability. Thus, although the big power plant might be able to meet some incremental demand in a spatially, temporally, and relationally narrow field, it cannot effortlessly be scaled up (or scaled back down) to resolve a national or global energy crisis. Linking big, scalability, and fragility we therefore advance the following:

> *Proposition 12. Fragility arises when big is forced into doing what was best left to the scalable.*

# FRAGILITY IN ACTION: EVIDENCE FROM LARGE DAMS

The 12 propositions about fragility above are based on a literature study. We now test the propositions against evidence from capital investments in 245 large dams. Our study of large dams began as an inductive enquiry into an under-researched question: will the brisk building boom of hydropower mega-dams underway from China to Brazil yield a positive return? Our previous effort (Ansar et al. 2014) found strong evidence that most large dams were too costly and took too long to build to deliver a positive risk-adjusted return. Our empirical results motivated us to begin theory development that would help us answer why investments in large dams might be particularly prone to what we above call "investment fragility," i.e. for the benefit-cost ratio to fall below 1?

Figure 3 presents an overview of the dataset by regional location of the dams, wall height, project type, vintage, and actual project cost. With a total value of $353bn (2010 prices) built between 1934 to 2007 on five continents in 65 different countries, our dataset on large dams is the largest of its kind. All large dams for which valid and reliable cost and schedule data could be found were included in the study. The data and the inductive study are described in Ansar et al. (2014).

**Figure 3: Sample Distribution Of 245 Large Dams (1934-2007), Across Five Continents, Worth USD 353B (2010 Prices)**

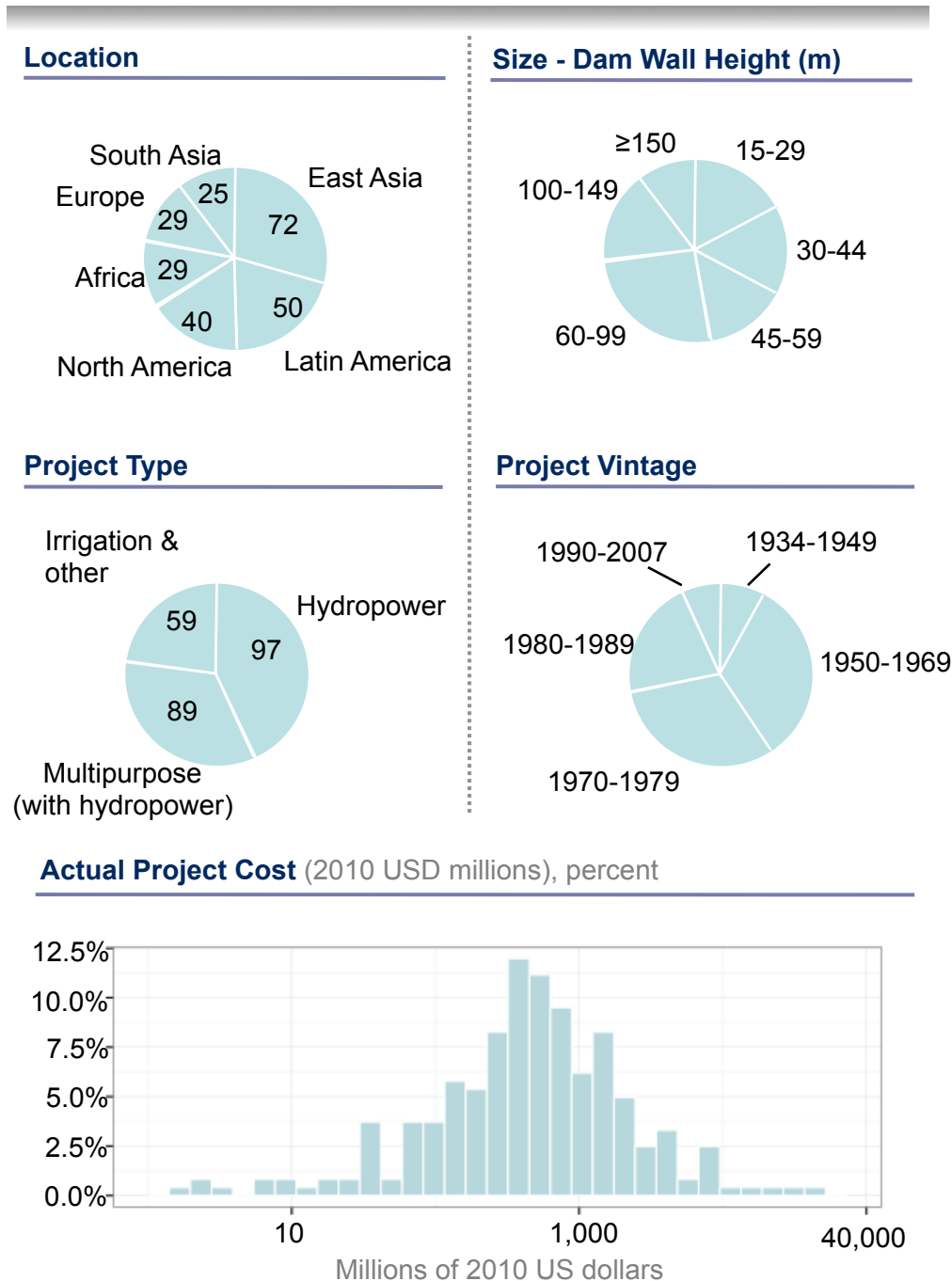

*Source: Ansar et al. (2014)*

**Why Study Big Dams?**
We find that large dams serve a particularly useful empirical setting to develop theory on investment fragility for the following reasons.

*First, big dams are archetypical megaprojects.* Dams are bespoke, site-specific constructions that have large physical proportions; require mobilization of vast quantities of inputs (materials, land, labor, or physical equipment); entail huge up-front financial outlays; generate a large number of unit of outputs; take a long time to build; last a long time; are spatially fixated; are highly complex not only in terms of number of constituent parts but also in terms of high interdependencies between constituent parts. Large dams typically also impact large numbers of people and the environment in their construction, operation, and eventual removal.

*Second, dams are indivisible and discrete assets.* A 90% complete dam is as valueless as a dam not built at all. For this reason, the physical scope and costs associated with a dam are methodologically well-defined and therefore particularly well-suited for like-for-like comparisons between planned and actual outcomes.

*Third, the main part of the whole-life costs of a dam are rendered up front during the construction phase.* Unlike, thermal power plants, dams do not require large variable costs in subsequent time periods for feedstock. There is a degree of transparency possible with the cash flows (or streams of costs and benefits) associated with dams that is much more difficult to establish for other big capital investments.

*Finally, results from large dams are generalizable to other big venture.* Large dams are a widely studied engineering problem, or what conventional theory might consider "a standardized production technology" (Sidak & Spulber, 1997), generating basic services, such as electricity or water, as outputs with seemingly established demand trends. Intuition would suggest that the overall planning uncertainties ought to be more limited in large dams than, for example, big capital investments based on revolutionary new technologies. A discovery of systemic errors in the building of dams would be indicative that the problems of investment fragility will be even more severe for less standardized capital investments.

**What Happens To Big Dams?**
Proposition 10 in the previous section suggested that if a system or process is systematically delivering poor outcomes it is an indicator of fragility. Specifically, we are interested in testing whether benefit-cost ratios fall below 1.

*Cost Performance[vii]*
With respect to cost overruns, we make the following observations:
- 3 out of every 4 large dams suffered a cost overrun in constant local currency terms.
- Actual costs were on average 96% higher than estimated costs (median 27%; *IQR* 0.86). The evidence is overwhelming that costs are

systematically biased towards underestimation and overrun (Mann-Whitney-Wilcoxon $U$ = 29646, $p$ < 0.01); the magnitude of cost underestimation is larger than the error of cost overestimation ($p$ < 0.01). The skew is towards adverse outcomes (i.e. going over budget).

**Graphing the dams' cost overruns reveals a long tail as shown in**
- Figure 4; the actual costs more than double for 2 out of every 10 large dams and more than triple for 1 out of every 10 dams. The long tail suggests that planners have difficulty in establishing probabilities of events that happen far into the future (Taleb, [2007] 2010, p. 284).

**Figure 4: Density Trace Of Actual/Estimated Cost (i.e. Costs Overruns) In Constant Local Currency Terms With The Median And Mean ($N$ = 245)**

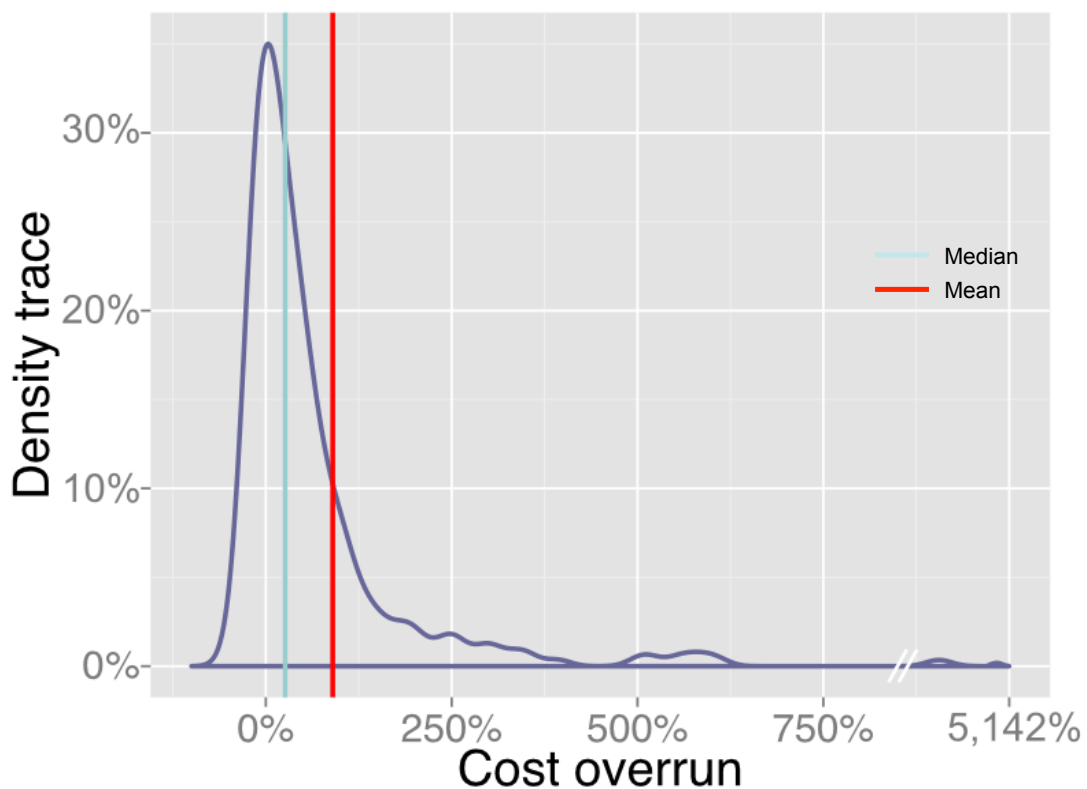

*Schedule Performance[viii]*
Not only are large dams costly and prone to systematic and severe budget overruns, they also take a long time to build. Large dams on average take 8.6 years. With respect to schedule slippage, we make the following observations:

- 8 out of every 10 large dams suffered a schedule overrun
- Actual implementation schedule was on average 44% (or 2.3 years) higher than the estimate (median 27%, or 1.7 years) as shown in Figure 5. A schedule overrun is detrimental to the benefit-cost ratio as it delays when much-needed benefits come on line. This decreases the net present value of future benefits. Even without a cost overrun, a schedule delay in and of itself can cause investment fragility. As the bulk of the costs of a big dam are incurred upfront they are not as sensitive to the discount rate as the benefits, which arrive farther in the future.

- Like cost overruns, the evidence is overwhelming that implementation schedules are systematically biased towards underestimation (Mann-Whitney-Wilcoxon $U$ = 29161, $p < 0.01$); the magnitude of schedule underestimation (i.e. schedule slippage) is larger than the error of schedule overestimation ($p < 0.01$).

**Graphing the dams' schedule overruns also reveals a long tail as shown in**
- Figure 5, albeit not as long as the tail of cost overruns. Costs are at a higher risk of spiraling out of control than schedules.

**Figure 5: Density Trace Of Schedule Slippage ($N$ = 239) With The Median And Mean**

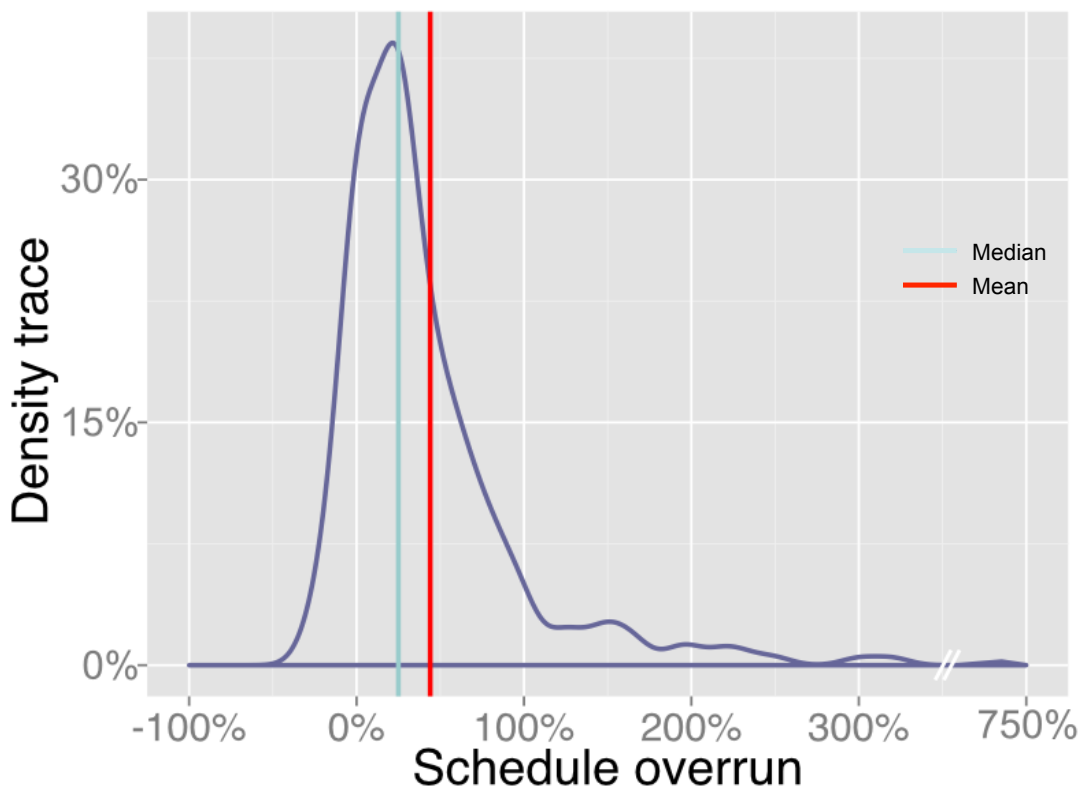

*Investment Fragility Threshold For Big Dams*

*Proposition 2* advanced that different systems have different fragility thresholds. For big dams the typical forecasted benefit-cost ratio (BCR) was 1.4. In other words, planners expected the net present benefits to exceed the net present costs by about 40%. With no change in future benefits or operations & management costs, a project suffering a cost overrun ratio of 1.4 or greater thus breach the fragility threshold—its BCR would fall below 1 and the broken asset's upfront sunk costs would become unrecoverable.

- The absolute threshold of 1.4 for big dams is broadly in line with many other physical infrastructure assets such as road, rail, bridge, or tunnel capital investments that too typically expect to generate a BCR of approximately 1.4 (NAO, 2014).

- Using the BCR of 1.4 as the fragility threshold, we found that investments in big dams are more fragile than investments in road, bridge, or tunnel projects. Investments in big rail projects are, however, even more fragile than investments in big dams. We arrived at this conclusion comparing the evidence on 245 big dams with a sample of another 353 roads, rail, and fixed link (bridges and tunnels) projects (see Flyvbjerg et al., 2002, 2003, 2005; Ansar et al., 2013).

- Investments in nearly half (47%) of the dams broke the fragility threshold, i.e., a cost overrun ratio of 1.4 or greater, compared to 14% for roads ($n = 239$, $M_{\text{Cost Ovverun}} = 1.22$) and 25% for fixed links ($n = 39$, $M_{\text{Cost Ovverun}} = 1.37$). For rail projects ($n = 79$, $M_{\text{Cost Ovverun}} = 1.44$) just over 50% of the sample had a cost overrun ratio greater than 1.4.

- In terms of runaway cost overruns, however, investments in big dams have a worse performance than rail projects. For big dams the 80th percentile cost overrun is nearly double the original estimate ($P80_{\text{Cost Ovverun}} = 1.99$) and the 90th percentile more than triple ($P90_{\text{Cost Ovverun}} = 3.07$) suggesting a long and fat tail. For rail the skew towards runaway cost overruns is less pronounced ($P80_{\text{Cost Ovverun}} = 1.75$ and $P90_{\text{Cost Ovverun}} = 1.90$) suggesting a truncated tail compared to big dams, though still long and fat compared to a normal distribution.

- Our calculations of the fragility threshold of big dams have a conservative bias. We have assumed that dams' benefits did not also fall short of targets, even though there is strong evidence in the existing literature that this is the case (WCD, 2000a; McCully, 2001; Scudder, 2005). We made this assumption because a comprehensive dataset of planned versus actual benefits of dams, like we have built for cost and schedule of big dams, does not yet exist. Data we have available on 84 hydroelectric large dam projects from our sample thus far suggests that they suffer a mean benefits shortfall of at least 11%.

Note that big dams suffer from the fundamental unfavorable asymmetry of fragile systems: the potential gain from big dams is capped but the pain is uncapped (see Figure 2). For example, in years of extreme drought a big hydropower dam may produce next to no power, but in years of floods the dams' ability to generate electricity is capped at the maximum of its design limit. So the maximum benefits the dams can produce are bounded but maximum losses—either from upfront cost overruns or subsequent operation & maintenance costs—are not. Factoring in benefit shortfalls and cost overruns on operation & maintenance costs, a bigger proportion of investments in big dams likely break than our conservative calculations suggest.

In summary, our conservative estimates suggest that investments in nearly half the dams break before the big dams even begin their operational life. This is due to outsized cost overruns on the upfront capital expenditure. Fragility at such an early stage of an investment's life cycle can be likened to an investment stillbirth. Subsequent discussion will show that the investment fragility risk only increases for durable and immobile big assets as their life progresses due to "cumulative fragility."

**Why Do Investments in Big Dams Break?**
The explanation for why investments in big dams break can be found in the thought processes that lead to their construction. We reviewed previously confidential business cases (called *Staff Appraisal Report* or *Project Appraisal Document*) at the World Bank presented to the Bank's Board prior to a final decision to build big dams. A typical business case puts forth the following line of argumentation to justify financing for a dam:

- First, the business case outlines the massive electricity or water shortfall a country faces;
- Second, the business case argues that the country possesses large untapped hydropower resources;
- Third, the business case proposes that a big dam should be built on a site deemed to be particularly advantageous by experts. The business case goes on to argue that the proposed big dam will be a big increment in solving the electricity or water scarcity of the country;
- Fourth, the business case cites technical studies which show that the big dam will be capable of *quickly* filling the gap between current market need and market supply *and* offer slack to meet surging future demand;
- Fifth, the business case also cites studies by economists, that the proposed big dam will be the "least costly expansion path that will adequately meet the projected demand" owing to its "economies of scale" (U.S. Department of Energy, 2009). Usually only one or two other alternatives are considered;

- Finally, detailed calculations are shown—sometimes spanning several appendices—that the present value of benefits of the proposed big dam, as of the date of the decision to build, will exceed the present value of the costs. In other words, that the benefit-cost ratio will exceed 1. Typically, little to no data on costs and benefits are shown for competing alternatives.

This thought process resembles more the rationalization of a pre-selected solution (a large dam) than the rational assessment of alternative ways to solve a given problem (provision of water and electricity). The process is rationalization presented as rationality (Flyvbjerg 1998). We identify several deep flaws.

First, although the typical business case correctly identifies the challenge as one of scalability – i.e., to *effortlessly* turn on or off the supply of electricity and water anywhere (*spatial)*, anytime (*temporal*), and for anyone (*relational*) – instead of proposing a scalable solution the business case proceeds to propose a single-shot big fix. Recall *Proposition 12*: *Fragility arises when big is forced into doing what was best left to the scalable.*

Second, the number of alternatives considered against the preferred solution is few if any. Alternatives such a decentralized generation or improving energy productivity are ignored. As Priemus (2008, p. 105) illustrates, it is typical for planners to settle early on a big asset as the preferred solution and then alternatives are "whittled down to nothing." This limits the option set presented to decision makers.

Third, by invoking the logic of purported economies of scale (by way of least cost expansion path), the business case then proposes building a staggeringly big dam since the bigger the dam the more scalable and efficient it appears on paper. Even as a typical business case worries a great deal about efficiency that might arise out of big size, it fails to dwell on the non-linear risks that will arise from big commitments to durable and immobile assets.

*The Various Facets Of Investment Fragility in Big Dams*
Thus far we have empirically substantiated the essential argument of this chapter that big capital investments are prone to fragility particularly when big is used as a blunt substitute for scalability. In particular we have looked at evidence for *Propositions 2, 10, and 13*. We now turn to finding empirical support for other propositions advanced in the theory section.

*Proposition 1* advanced that fragility is typically irreversible. In the empirical setting of investments in big dams this proposition takes on two facets. First, once the investment is broken – typically due to an upfront cost overrun – it is nearly impossible to reposition the asset to generate a positive return—i.e. increasing revenues to pay back the escalated up-front costs and debt financing on them proves elusive as best. The Yacyretá Hydroelectric Project—a joint venture between Argentina and Paraguay under the Yacyretá Treaty on the Parana River started in 1973 and only fully completed 38 years later in 2011 according the treaty's intended proposal—is a telling case. The dam operated at 60% of its intended capacity from 1998 onwards. In June 2001, the World Bank—one of the dam's financiers—published its *Implementation Completion Report* (ICR), which arrived at the following conclusion about the project's outcome:

> Based on the above, it is clear that the project has not met its goals as there is no solution on how to terminate the project, and operate at full capacity. As explained, the project has operated for a prolonged period of time at an unintended level [below intended capacity]. In addition, the project presents a poor ERR [Economic Rate of Return], negative NPV [Net Present Value] recalculated by this ICR, and uncertain project sustainability. As a result, the outcome of Loan 3520-AR is rated unsatisfactory, and the outcome of the Yacyretá scheme as a whole is rated unsatisfactory mainly because of its big negative NPV ( p. 6).

The project's lower-than-expected volume of output was exacerbated by a lower-than-expected price of electricity. The World Bank (2001) report goes on to explain that the Yacyretá dam has a "big negative NPV" because "the spot price [of electricity sold] has been (and is estimated to remain so after year 2001) lower than US$30 per MWh [agreed in the 40-year 1973 treaty]." The ICR did not envision a scenario in which the dam would revert to a positive NPV due to the sheer scale of costs sunk into the project.

Second, the breaking of an investment—unlike the porcelain cup at the beginning of this chapter—happens in slow motion. Yet decision-makers appear unable to reverse commitment to the losing course of action. In the case of Yacyretá dam, the World Bank's *Project Performance Audit Report*— prepared by the World Bank's independent evaluation department, found: "[World] Bank performance is rated as unsatisfactory. The appraisal overlooked the downside risk of a slower demand growth, which proved to be of paramount importance. Later, opportunities to reassess the project were wasted in spite of the recommendations made by a good quality and timely *Public Investment Review (PIR)* in 1985. The Bank also accepted non-compliance with financial covenants" (World Bank, 1996b, p. 14). The Yacyretá dam's *Project Performance Audit Report* (World Bank, 1996b, p. 16) neatly summed up the insidious but irreversible fragility of big capital investments, "In several occasions the Bank had a good case for stopping the project before the major civil works were too advanced." But it did not. The phenomenon, whereby smart people and organizations are unable to stop themselves from throwing good money after bad goes by many names in the academic literature such as escalation of commitment or lock-in (McNamara et al., 2002). The financial magnitude of a big venture is so large that once started the commitment turns into a binding, ruinous, co-dependence.

*Proposition 3* suggested that proportionality matters when considering the impact of fragility. In the case of dams, an investment in a big dam built in a poor country plays out very differently when compared to an investment in a similar big dam built in a rich country. This position appears so obvious as to be trivial, but the real-life effects are striking.

Big dams built in North America (*n* = 40) have considerably lower cost overrun (*M* = 11%) than big dams built elsewhere (*M* = 104%) as show in Figure 6. Note, however, that 3 out of 4 dams in our reference class had a North American firm advising on the engineering and economic forecasts. Consistent with anchoring theories in psychology, we interpret this observation as an indication of an overreliance on the North American experience with large dams, which appears to be biasing cost estimates downwards in the rest of the world. Experts may be "anchoring" their forecasts in familiar cases from North America and applying insufficient "adjustments" (Flyvbjerg et al., 2009; Tversky and Kahneman, 1974), for example to adequately reflect the risk of a local currency depreciation or the quality of local project management teams. Proponents often hold up the Hoover Dam—finished in 1936 in the U.S.—as a prime example for why similarly big dams ought to be built in poor countries. Evidence shows, however, that the likes of Hoover dam are false analogies. Instead decision makers in developing countries should consider evidence for the entire dam record, and in the case of big dams, investment fragility is the most likely outcome in developing countries.

**Figure 6: Location Of Large Dams In The Sample And Cost Overruns By Geography**

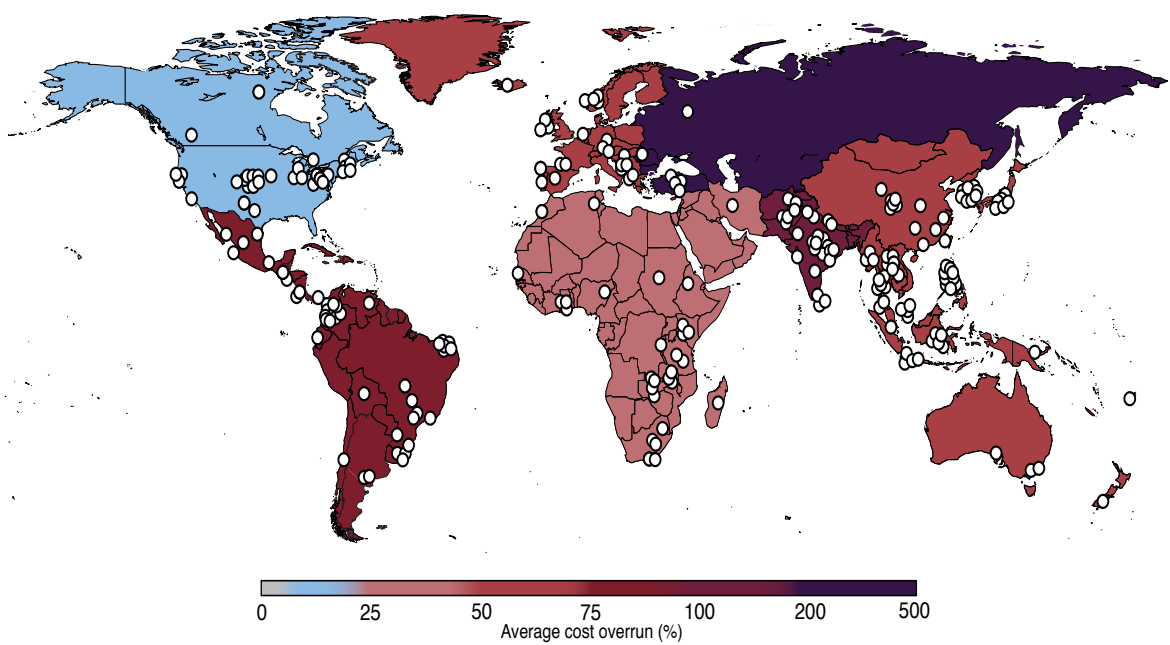

Similarly, with respect to construction schedule, we find that countries with a higher per capita income tend to have lower schedule overruns than countries with lower per capita income. We concur with the interpretation of Bacon and Besant-Jones (1998, p. 325) that "the best available proxy for most countries is [the] country-per-capita income…[for] the general level of economic support that a country can provide for the construction of complex facilities". This result reinforces the argument that developing countries in particular, despite seemingly the most in need of complex facilities such as large dams, ought to stay away from bites bigger than they can chew.

Finally, when an investment in a big dam breaks in a poor country, the country's economy as a whole more readily inherits that fragility by way of accumulating debt. Brazil's Itaipu Dam was built in the 1970s. It cost nearly $20 billion, 240% more in real terms than predicted and it impaired Brazil's public finances for three decades. Similarly, the actual cost of Tarbela dam, the majority of which was borrowed from external sources, amounted to 23% of the increase in Pakistan's external public debt stock between 1968-1984; or 12% for Colombia's Chivor dam (1970-1977) as shown in Table 1. Companies and countries with insufficiently large balance sheets to absorb adverse outcomes of big bets gone awry face financial ruin.

**Table 1: Total Stock Of Public Net External Debt (US$ Current, MM)**

| Year | Colombia | Pakistan |
|---|---|---|
| 1968 | | 3,252.4 |
| 1970 | 1,296.6 | |
| 1977 | 2,699.6 | |
| 1984 | | 9,692.8 |
| Debt increase over the implementation schedule | 1,403.0 | 6,440.5 |
| Cost of mega-dam over the relevant period (US$ current MM) | Chivor dam | Tarbela dam |
| | 168.7 | 1,497.90 |
| Cost of dam as percentage of debt increase | | |
| | 12.0% | 23.2% |

*Source: Ansar et al. (2014)*

As the discussion leading to *Propositions 5 & 6* above showed, there is typically an aspiration to make a seemingly fragile system more robust. In investment fragility terms, this means building in a contingency as a buffer against cost overruns. The conventional logic being that if a project returns a BCR greater than 1 even after taking the contingency into account then the investment can be considered robust—i.e. it will not easily cross the fragility threshold. Evidence suggests, however, that standard contingencies are too low. For example, in providing a contingency against inflation for big dams in our sample, forecasters expected the annual inflation rate to be 2.5% on average, but it turned out to be 18.9% (averages for the entire sample). Had adequate contingencies been provided, the benefit-cost ratios would have been lower and investments would not have readily received the final go-ahead.

Sensitivity analysis is another mechanism by which an aspiration to robustness is formalized in business cases of big capital investments. But like contingencies, this exercise—as customarily conducted in business cases of big capital investments—is perfunctory. A typical *ex ante* sensitivity analysis stress tests for a very narrow range of variance around the base case, whereas the actual *ex post* evidence shows that a very wide range ought to have been tested. The difficulty is that when the sensitivity analysis range is widened from the narrowest confines to reflect real-life variance, the business case falls apart. The before versus after picture of Colombia's Guavio hydroelectric project is illustrative.

Figure 7 exhibits a facsimile of a representative section on benefit-to-cost analysis from the World Bank's 1981 *Staff Appraisal Report* of Colombia's Guavio hydroelectric project, which gave the final decision to build the dam. As the facsimile illustrates, planners of Guavio first conducted a base case scenario that showed that the dam had a positive economic rate of return (of 15%) greater than the opportunity cost of capital (11%). Planners then conducted a sensitivity analysis on the base case benefits and costs but only subjected them to a stress test range of ±15%. The planners, despite acknowledging that their estimated were subject to uncertainty, appear reluctant to test for extreme events even if such events are characteristic for large dams, as we saw with the long, fat tails above.

**Figure 7: An Aspiration To Investment Robustness in Guavio Hydroelectric Project**

6.04    On this basis, the return on investment is about 15.0% (Annex 6.1), which compares favorably with the opportunity cost of capital for Colombia, estimated to be 11%.

6.05    A sensitivity analysis was carried out to estimate the impact on the internal financial return of possible changes in cost and revenues. These are summarized below:

Rate of Return on Investment Program (%)

| Cost \ Benefits | 85% | 100% | 11% |
|---|---|---|---|
| 100% | 11 | 15 | 17 |
| 115% | 10 | 12 | 15 |

If the cost of the program increases by 15%, the rate of return would be about 12%. If the program costs do not increase but benefits are 15% lower than estimated, the return would be about 11%. If the program costs increase by 15% and the benefits decrease by 15% the return on the investment program would be about 10%. The equalizing rates obtained understate the real economic rate of return of EEEB's investment program since revenues from electricity sales do not fully reflect all benefits to society.

*Source: World Bank (1981, p. 42)*

As it turned out, the Colombian utility that owned the Guavio dam went bankrupt (World Bank, 1996a, p. iii). The dam's actual outturn costs (in local currency constant prices) spiraled six times their estimate. Instead of the +15% cost overrun that the *ex ante* sensitivity analysis allowed for, a +500% stress test would have been more appropriate! The investment in Guavio did not just break. It shattered.

The World Bank's *Implementation Completion Report*, which reported the *ex post* outcomes of the Guavio dam conceded that the investment in Guavio was broken even before the hydroelectric project began operational life. "Generally speaking the project's outcome has to be considered highly unsatisfactory…this has been at a high cost, particularly in terms of the non-viability (financial and economic) of the project at completion" (World Bank, 1996, p. iv).

*Proposition 7* advanced that cumulative, and often hidden, processes increase a system's fragility over time. With respect to big dams, we have thus far focused on investment fragility that happens at the beginning of the life of a big dam due to a cost overrun that exceeds the expected future benefits. Now we turn to the "cumulative investment fragility" of big dams as they age. This occurs when large costs are incurred to repair, overhaul, or decommission the ageing big dam. These late-life costs can turn an aging big dam into a liability in perpetuity. We review two case examples to elucidate cumulative fragility both from a material and an investment perspective.

Cumulative material fatigue from minor stressors, poor maintenance, and new demands placed on an infrastructure can cause physical collapse without warning. The sudden structural collapse of the Stava dams in the Eastern Italian Alps on 19 July 1985 that resulted in 268 human fatalities is a tragic case of physical fragility brought on due to years of neglect. The Stava dams, even though visibly in a state of decay, had kept on performing their intended function. Since the cost of rehabilitating the dams was high, the owners—the Prealpi Mining Co. of Bergamo[ix]—felt little pressing need to act. The dams then collapsed with no apparent discreet cause—a hallmark of cumulative fragility (see Luino & Graff, 2012, p. 1042). The physical fragility of the dams and the resulting casualties also caused investment fragility: financial liabilities for the clean up exceeded €130 million and criminal sentences for many of the individuals involved in the operation of the dams.

A dramatic physical collapse is, however, not necessary to inflict cumulative investment fragility. The Kariba dam, on the border between Zimbabwe and Zambia, presents a salient and current case[x]. The Kariba hydroelectric program was built in two phases. The first phase entailed the construction of the dam wall, which had a height of 128 m and impounded Lake Kariba—the largest man-made lake in the world at the time. In contrast to the typical big dam, Kariba Phase 1 was built at breakneck speed. The civil works were completed in just four years from 1955 to 1959. The first electricity was generated in January 1960 (WCD, 2000b, p. 9). All 6 generators installed during the first phase were in operation by 1962, with a capacity of 666MW—which exceeded the intended scope of 500MW. Kariba's Phase 1 is also among the rare big dams that came in on budget; built within its then £80.8 million budget envelope (World Bank, 1956; WCD, 2000b).

Kariba's Phase 1 thus appears to have beaten the odds to a healthy birth. But cumulative processes have pushed Kariba towards investment fragility[xi]. Outflows of water from the dam's sluice gates have, over time, eroded the plunge pool below the gates. The pool, which was initially 10m deep has gradually worn down to over 80m eroding towards the dam wall's foundations as shown in Figure 8. The structural integrity of Kariba dam is now under a pressing threat.

**Figure 8: Cumulative Fragility: Erosion Near The Foundation Of Kariba Dam Wall**

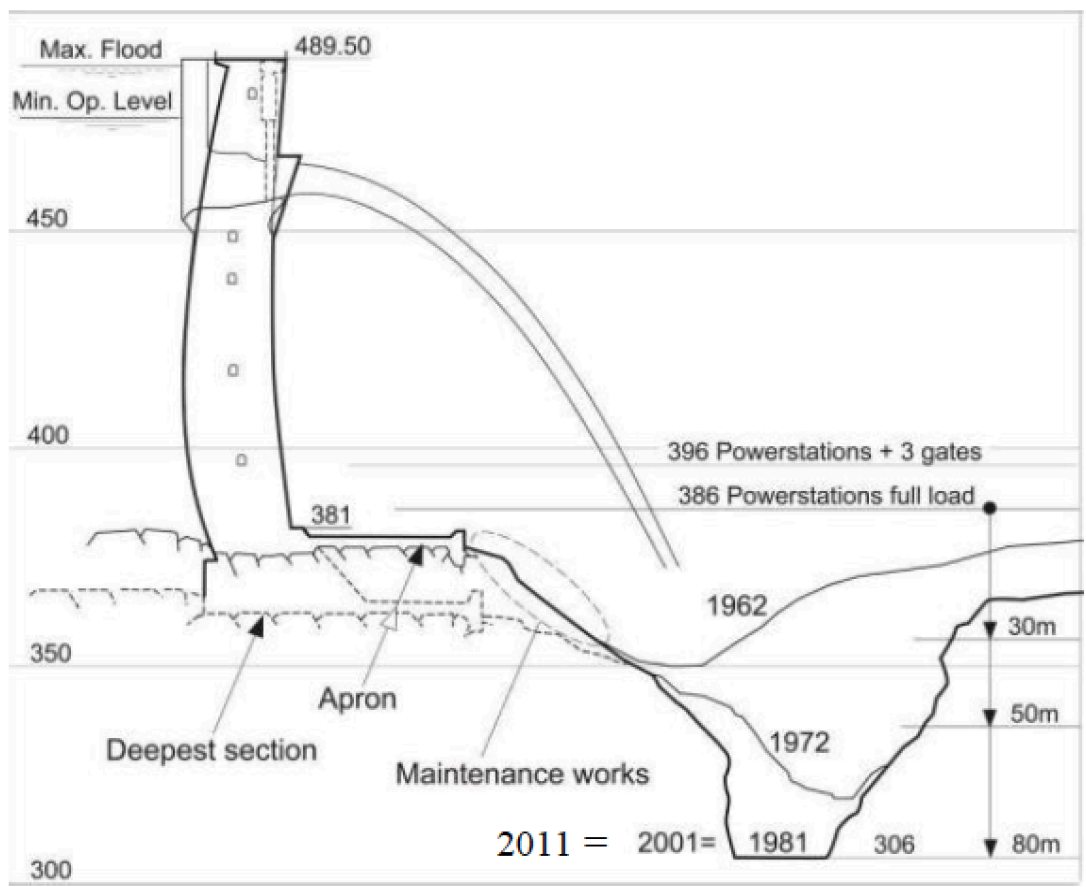

*Source: World Bank (2014, Figure 2)*

The cumulative fatigue of Kariba dam's structure has placed decision makers in an unenviable "squeeze": either allow the dam to collapse and bear unthinkable human, environmental, and financial losses or embrace huge outlays to maintain the dam in perpetuity. For now four institutions—the European Union, Swedish Government, African Development Bank, and the World Bank—have spurned into action to provide $295 million in loans to undertake repairs that are expected to take at least ten years (World Bank, 2014). Although the scour in the plunge pool is the most pressing problem facing the Kariba, a number of other issues have begun to creep up such as the swelling of the concrete and malfunctioning of flood gates. Kariba's maintenance bills are unlikely to be a fixed predictable sum with a clear end date. Whether donors will find sources of finance in perpetuity to maintain the dam is an open question.

Kariba's material fatigue is a reminder that big dams have finite life spans. Even if not in the next decade or two, in foreseeable human history a big dam like Kariba will have to be removed (even if it were to be rebuilt) lest it were to accidentally collapse. No on has the remotest idea how much will need to be spent on the end-of-life arrangements of a big dam like Kariba. Such rehabilitation and demise costs were not anticipated in Kariba's original BCR analysis. 95% of the world's big dams were built in the last century and many are approaching the stated end of their service lives (Lavigne, 2005). Will dams less illustrious than the Kariba find resources for repairs in perpetuity? Or will the forgotten dams only be heard of when they collapse from cumulative fatigue like the Stava dams? There are over 45,000 big dams in the world. Who will pay for their decommissioning or reconstruction?

*Proposition 8* suggested that cognitive biases make it difficult for laypeople and experts alike to detect fragility. The Vajont dam in Italy is perhaps a particularly unfortunate case illustrative of psychological and sociological dimensions of fragility. Located 100 km north of Venice, Vajont was the world's highest thin arch dam with a height exceeding 260 meters—equivalent to a 60-storey skyscraper—at its completion in 1960. The dam was built across the Vajont Valley, a deep, narrow gorge characterized by massive, near-vertical cliffs. On 9 October 1963 a giant landslide collapsed into the reservoir at an unanticipated speed, which generated a 250-meter tsunami that overtopped the dam destroying villages downstream. Nearly 2,000 people died. Apart from levying exacting human loss, the dam also lost its economic function.

Although the proximate cause of Vajont's fragility appears to be an unexpected natural disaster, the root causes point to biases of human judgment under uncertainty (Tversky and Kahneman, 1974; Kahneman and Lovallo, 1993; Lovallo and Kahneman, 2003; Kahneman, 2011). The literature in psychology suggests that experts (e.g., statisticians, engineers, or economists) and laypersons are systematically and predictably prone to errors when forming judgments under uncertainty. Biases, such as overconfidence or overreliance on heuristics (rules-of-thumb) underpin these errors, which result in the underestimation of risk.

The behavior of the experts responsible for building Vajont is consistent with such cognitive biases: By November 1960 it had become clear that the rock mass sloping towards the reservoir behind the dam was severely fractured. Experts reached the conclusion that it was not possible to completely stop the landslide but nonetheless "assumed that by elevating the level of the reservoir in a careful manner...the rate of movement [of the sliding mass] could be controlled" (Petley, 2008; also see Semenza & Ghirotti, 2000; Kilburn & Petley, 2003). Experts' psychological confidence that they could control the landslide is consistent with the notion of "illusion of control," a tendency for people to overestimate their ability to control outcomes (Kahneman & Riepe, 1998). Experts' illusion of control was exacerbated by a systematic underestimation of variance of the underlying risk factors they faced. Experts built several scale models to simulate the potential outcomes of a landslide into the reservoir behind the Vajont dam wall. However, these models did not consider the cumulative effect that a large mass might have if it slid at very high speed, which turned out to be the eventual outcome (Marco, 2012, p. 145-46).

**In line with *Proposition 9*, we find that at the organizational level such psychological biases appear not to be readily attenuated. For example, we analyzed whether cost estimates of big dams have become more accurate over time. The idea being that even if psychological causes led to errors in earlier projects, organizational factors would intervene over time to attenuate psychological biases. In contrast, our statistical analysis suggested that irrespective of the year or decade in which a dam was built there are no significant differences in forecasting errors ($F = 0.57$, $p = 0.78$). Similarly, we did not observe a trend indicating improvement or deterioration of forecasting errors ($F = 0.54$, $p = 0.46$) as also suggested by**
Figure 9. Organizations appear to not be learning from past mistakes. Why organizations fail to attenuate adverse outcomes of psychological biases continues to be a lively area of future research (see Durand, 2003; Makadok, 2011).

**Figure 9: Inaccuracy Of Cost Estimates (Local Currencies, Constant Prices) For Large Dams Over Time (N=245), 1934-2007**

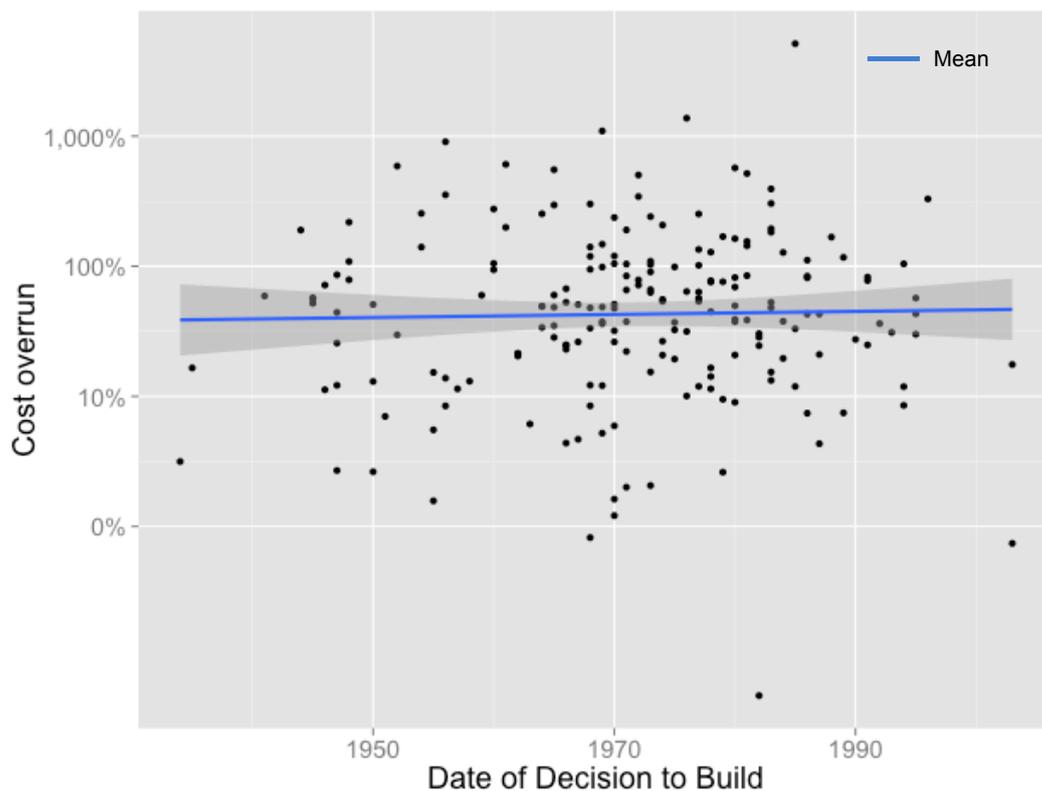

# SUMMARY AND CONCLUSIONS

Megaprojects, in the private and public sectors alike, are more ubiquitous and bigger than ever before. These big capital investments are justified on the basis of theories of big that underpin two aspirations: An aspiration to efficiency from purported economies of size or scope; and an aspiration to scalability from building capacity ahead of demand. We found that big investments have a poor track record in delivering on either aspiration. Contrary to the conventional proposition of "bigger is better," we found big capital investments to have a propensity to fragility—vulnerability of the investments to become unrecoverable due to the impact of random events(s).

Our theoretical argument can be summarized as follows: the emphasis of theories of big on efficiency and building capacity ahead of demand encourages managers to take "definitive, static bets" on big ventures (Klingebiel &Adner, 2015, p. 222). What theories of big fail to incorporate in their logic is that oversizing a system increases its complexity disproportionately due to the greater number of permutations of interactions now possible among more sub-components, and this leads to fragility. Small errors in one or a few interactions of the bigger system magnify. All of this exerts a financial toll: propensity to cost overruns at construction; unexpectedly large bills in fixing new vulnerabilities when they become apparent as the system ages; and huge clean-up costs if the system has to be decommissioned or if it collapses. Despite their Goliath appearance, big capital investments break easily—i.e. deliver negative net present value-- when faced with disturbances. A greater propensity to fragility is intrinsic to big investments.

We further argue that conventional theory has conflated big with scalability. Big entails being bespoke and complex with multiple attributes such as large upfront financial outlays, durable temporal horizons, spatial immobility, and ability to produce large units of outputs. In contrast, scalability is best understood by considering the difference between a live musical performance and a digital recording of the same music. The live concert is only available at the prescribed place (spatial), at the prescribed time (temporal), and to a limited audience (relational). Anyone can enjoy the digital recording, anywhere, anytime. The live concert does not scale. The recording is scalable. Fragility arises when big is forced into doing what was best left to the scalable.

We substantiate our argument using evidence from a dataset of 245 big dams with a total value of $353bn (2010 prices) built 1934 to 2007 on five continents in 65 different countries. Our primary findings are:
- Nearly half the dams we studied suffered a cost overrun so large for the projects to be considered broken even before they began operations. That is, the capital sunk up front could not be recovered. Fragility at such an

- early stage of an investment's life cycle can be likened to an investment stillbirth.
- Cost risks for big dam have fat tails – the actual costs more than doubles for 2 out of 10 dams; triples for 1 out of every 10 big dams.
- Managers do not seem to learn. Forecasts today are likely to be as wrong as they were between 1934-2007.
- Costs aside, big dams take a very long time to build, on average 8 years, which is a late coming to solve pressing energy and water needs.
- Investment fragility risk increases as the life of a big dam progresses due to cumulative processes—such as material fatigue and threat of catastrophic failure—which trigger need for costly rehabilitation in perpetuity.
- Costs of maintaining or removing big dams past their intended service lives is a looming financial liability.
- The companies or countries that undertake big investments inherit their fragility. For example, the outsized costs of big dams and similar megaprojects have caused an explosive growth of debt in developing countries hamstringing their economic potential. Developing countries, despite seemingly the most in need of complex facilities such as big dams, ought to stay away from projects that are big relative to the national economy.

Ought decision-makers abandon all big ventures? Of course not. But decision-makers must carefully assess when bigger is better, instead of unthinkingly assume this is the case. Evidence suggests that on a risk-adjusted basis more often than is assumed big ventures are unlikely to present good value. Top decision-makers responsible for giving the final green light to a big capital investment ought to remain skeptical of the numbers presented to them at appraisal. Decision-makers should also seek to de-bias estimates of time to task completion, costs, and benefits by demanding more extreme stress tests, reflecting the full variance of phenomena, to determine the fragility threshold of the investment they are about to undertake. We would also advise against making big capital investments decisions on razor-thin margins. Big investments need far more cushioning to avoid fragility than current management practice tends to assume.

If big is fragile, what's the alternative? We do not have the space here to answer this question. Elsewhere, we study managers who work by the heuristic, "If big is fragile, then break it down," to deliver their projects successfully, and we study theories that support this approach, such as theories of modularization (Baldwin & Clark, 2000; Schilling & Steensma, 2001; Garud et al., 2003; Levinson, 2006); real options in product innovation (Klingebiel & Adner 2015), and greater user-producer co-development (Grabher et al., 2008, Ansar, 2012). We hope to report from this work later. For now, we conclude that prior to committing to a big venture, managers should rigorously consider possible alternatives. Scholars should similarly carefully re-evaluate theories of big. Our data indicate that the current tendency to accept theories of big at face value and lock in early in capital project decision making to a big "favorite solution," often erring towards the monumental, drives fragility and therefore incurs a high risk of failure.

# ENDNOTES

[i] Future efforts may require drawing more careful distinctions among the members of the disorder family (also see Goldstein & Taleb, 2007 on volatility).

[ii] Buldyrev et al. (2010, p. 1025) report, "For an isolated single network, a significant number of the network nodes must be randomly removed before the network breaks down. However, when taking into account the dependencies between the networks, removal of only a small fraction of nodes can result in the complete fragmentation of the entire system."

[iii] Not all things fit neatly in the proposed 2 X 2 matrix in Figure 1. Although it is possible to glue back together a broken antique porcelain cup to restore the function of drinking tea, it is aesthetically displeasing. Similarly, Humpty the cannon could have been restored to its defensive function. But that proved unachievable in the time available despite the effort of "all the king's horses and all the king's men." The temporary fragility of the cannon, however, irreversibly spread to the Royalists forces. Thus even where a broken system is highly recoverable three questions still remain relevant: i) What effort (e.g., cost or time) will the recovery take? ii) What harm might occur to other things (i.e. inherited fragility) during the down time in which the recovery takes place? iii) Will the recovery yield a functional inferior? What this suggests is that to make a system that has a high fragility threshold and is easily restored to its original function is incredibly difficult.

[iv] In Greek mythology, when Achilles was a baby, it was foretold that he would die young. To prevent his death, his mother Thetis took Achilles to the River Styx, which was supposed to offer powers of invulnerability, and dipped his body into the water. But as Thetis held Achilles by the heel, his heel was not washed over by the water of the magical river.

[v] Other academic fields such as geography and ecology have also taken a keen interest in scale and typically use similar language—e.g., finer or coarse/broader scale—as Mandelbrot (Wiens, 1989; Wu & David, 2002). Perhaps the noteworthy difference is that whereas geometry tends to treat scale as a continuum, scholars in geography or ecology have preferred to conceptualize scale as a nested hierarchical structure such as Russian dolls. For example, a street, block, neighborhood, borough, city, province, country, region, continent, the planet, the solar system etc. are a nested spatial structure. Although, the nested hierarchical structure conceptualization a blunt approximation, as long as the zoomed-out scales in a nested structure are not conflated with bigger we are not against the use of such an approximation to aid understanding.

[vi] Our definition of scalability differs from the more popular conceptualizations, represented, for example, by Weinstock and Goodenough (2006). The differences are that Weinstock and Goodenough (2006) mainly focus on (1) First, relational scale concerns such as number of users or user driven performance metric, whereas space and time cannot be ignored; (2) Second, they focus almost entirely on the ability to scale up. Effortless scaling down is equally important, particularly when previous capacity becomes obsolete and needs to be removed. Thus it is not just important to meet increasing incremental demand but to know what to do if the demand disappears.

[vii] Actual outturn costs (also called actual CAPEX) are defined as real, accounted construction costs determined at the time of project completion. Estimated costs are defined as budgeted, or forecasted, construction costs at the time of decision to build. The actual outturn costs comprise the following elements: right-of-way acquisition and resettlement; design engineering and project management services; construction of all civil works and facilities related to completing a dam project; equipment purchases.

Cost underrun or overrun is the actual outturn costs expressed as a ratio of estimated costs. The year of the date of the decision to build a project is the base year of prices in which all estimated and actual constant costs have been expressed in real (i.e. with the effects of inflation removed) local currency terms of the country in which the project is located. Cost overruns can also be expressed as the actual outturn costs minus estimated costs in percent of estimated costs.

[viii] Actual implementation schedule of the project in months. The implementation start date (also known as the final decision to build) is the date of project approval by the main financiers and the key decision makers. The implementation completion date is the date of full commercial operation. Schedule slippage or schedule underrun or overrun is the actual implementation months express as a ratio of the estimated implementation schedule.

[ix] According to the New York Times (22 July 1985), The Montecatini Company had built the mine and Stava dam complex in the early 1960's. It was taken over by the Italian state energy company, Enil, in 1979 and sold to Prealpi in 1981. Disaster struck in 1985.

[x] Discussion of the Kariba case here benefited from personal communication with Jacques Leslie.

[xi] As part of its cumulative fragility, it is worth noting Kariba dam's large and continuing negative social and environmental impacts, which were poorly mitigated. Thayer Scudder (2005, p. 1) writes, initially "considered a successful project even by affected people based on cost benefit analysis, Kariba also involved unacceptable environmental and social impacts. The involuntary resettlement of 57,000 people within the reservoir basin and immediately downstream from the dam was responsible for serious environmental degradation which was one of a number of factors that left a majority of those resettled impoverished." Moreover, as an important aside, the Phase 2 of the Kariba built 1970-1977, constituting a 600 MW powerhouse, suffered a cost overrun 2.5 times its estimate badly damaging the overall program's BCR (see World Ban, 1983; WCD, 2000b, pp. 12-14).